# To study the structural, electronic and optical properties of predicted stable halide perovskites ABX$_3$

Kashif Murad, G. Murtaza, Muhammad Noman, Shamim Khan

Materials Modelling Lab, Department of Physics Islamia College Peshawar. KP, Pakistan

## ABSTRACT

Structure, optical and electronic properties of predicted stables Perovskites ABX$_3$ are calculated using DFT approach. The modified Becke-Johnson potential is also used to investigate electrical and optical properties. The density of states and electronic band structure calculations reveal that the predicted stable halides Perovskites ABX$_3$ have a direct as well as indirect band gap, with dielectric function, optical reflectivity, absorption coefficient, optical conductivity, extinction coefficient, refractive index are calculated in different ranges of energy. The maximum value of static dielectric function is observed for KNiI$_3$ and RbIrBr$_3$ are 3.5 and 3 respectively. The highest absorption peak among the all is observed at 26eV for SrLiF$_3$, while refractive index of that SrLiF$_3$ gives peak at 24eV then refractive index decreases below the unity as energy increases up to 27.5eV. The maximum peaks of optical conductivity of predicted stables halides Perovskites are observed in visible to ultraviolet range (1.7eV to 30eV). The maximum peaks of optical conductivity are observed for SrLiF$_3$ at 24eV. For the first time, all of the properties evaluated for predicted stable halides ABX$_3$Perovskites are reported, because no theoretical work is reported yet on most of the physical properties.

# 1. Introduction

Perovskites make up a large amount of the mantle of the earth crust, hence studying their physical characteristics is essential. Perovskites are a broad class of compounds that include anything from insulators to superconductors to colossal magneto resistive compounds. They exhibit a variety of interesting features; thermoelectricity, superconductivity, spins dependent transport, colossal magnetoresistance, charge ordering, and the interaction of structural, optical characteristics and magnetic, are all common aspects of perovskite materials. The perovskite materials are often utilized as sensors, catalytic electrodes in fuel cells, substrates, and they also considered interesting for optoelectronic prospects [1, 2].

Perovskites come in a variety of shapes and sizes, such as metal-based halides Perovskites, which have visible and near-infrared band gaps (and thus are more suitable for use in light emitting devices and also as a photovoltaic), with a big set of transition metal fluoride perovskites in the first row, which can be used as a host material for up or down conversion emission or for lithium intercalation [3-5]. $KMgF_3$ (parascandolaite) one of these fluoride perovskites [6], is also one of a pure cubic HPs perovskite structure discovered in nature. As a result, $KMgF_3$ is a better high symmetry crystallographic structure type than $SrTiO_3$ for HPs.

Because of their superior charge-transport and optical characteristics, halide Perovskites($ABX_3$) have gained a lot of interest in the scientific community as light harvesting materials, such is conformable band gaps, high external quantum efficiency, long charge carrier diffusion lengths, high chemical defect tolerance and high absorption coefficient. Due to their high quantum yield, multicolor luminescence and simple cheap growth the halide Perovskites become very famous [7], in short time, power conversion efficiency (PCE) of comparable photovoltaic cells improved from 3.81 percent to 23.3 percent. In addition to solar cells, the applications of HPs mainly include photo detectors (PDs), light emitting diodes (LEDs), encryption devices and lasers.

The computer neural networks have reported a list of the best stable halides Perovskites($ABX_3$) {$SrLiF_3$, $TlSrF_3$, $ZnScF_3$, $ZnLiF_3$, $TlBeF_3$, $CsBeCl_3$, $CaCuCl_3$, $CsRhCl_3$, $CsRuCl_3$, $BaAgCl_3$, $RbRhBr_3$, $RbPdBr_3$, $CaCuBr_3$, $BaAgBr_3$, $RbIrBr_3$, $CaCuI_3$,

KNiI$_3$, CsTcI$_3$, KIrI$_3$, CaAgI$_3$}. The newly predicted super symmetric perovskites still not being synthesize however some of the predicted ones e.g. ZnScF$_3$, ZnLiF$_3$, SrLiF$_3$, TlSrF$_3$, RbPdCl$_3$, CsBeCl$_3$ and TlBeF$_3$ were already describes in former theoretical works [8]. More to further understand the characteristics of these molecules, more experimental study is required. By knowing the properties and structure of these molecules theoretically makes the experimental work more productive.

The major goal of this study is to cover the lack of theoretical data on the optical properties of recently predicted stable halide perovskites. Since no theoretical work is reported yet on most of the perovskites and also the latest artificial intelligence algorithms predict the stable new perovskites. It would be very interesting to determine the optical properties of these newly predicted halide perovskites. I select these compounds and studied them using the density functional theory approach (DFT).The full potential linearized augmented plane wave (FP-LAPW) technique is used to calculate the parameters related to optical characteristics, with various exchange and correlation approximations. In terms of structural properties, the optimized lattice constants, ground state energies, pressure derivative of the bulk moduliand bond nature among the anions and cationshave all been investigated. Density of states and band structure are used to explain the electronic properties of these substances. Electronic charge density graphs represent chemical bonding. The complex dielectric function, optical conductivity, effective number of electrons, complex refractive index, energy loss function, absorption coefficient and reflectivity participating in optical transition are used to describe optical properties. This research will contribute to the knowledge of the optical characteristics of halide Perovskites and will also explain different important applications of these compounds.

Theory and Computational Details:

## 3.1. Density Functional Theory

Density Functional Theory (DFT) is a quantum mechanical method to treat the many body problems. Density functional theory is used to study ground state electronic structure of many-particle systems such as molecules, atoms, alloys, condensed phases, chemistry, physics, and materials science, which is the most efficient and consistent computational quantum mechanical approach. The electron density functional is used in DFT to show the specific distribution of electrons in a crystal or a material; as a result, the name density functional theory is derived from the electron density functional. Instead of the wave function, electronic density is used as an essential issue in density functional theory.

We found that, because to the restrictions of the Hohenberg and Kohn theorems, electron density can be utilized as a potential for the solving of several particle systems and that there is no other method to get around the precise charge density. Kohn and sham studied at the subject of non-interacting electron systems [9]. The wave function of a single particle can be used to calculate a system's kinetic energy. The effective single particle equation, which provides the density of the real interacting system, may be used to determine the wave function of the particle in the Kohn-Sham equation. The Kohn-Sham equation in the atomic system can be expressed as:

$$\epsilon_i \Psi_i = \left[\frac{-1}{2}\nabla^2 + V_{ett}(r)\right]\Psi_i$$

$$V_{eff}(r) = V_{ext}(r) + \int \frac{\rho(r)}{|r-\acute{r}|} d\acute{r} + V_{XC}$$

$V_{eff}(r)$ is the effective potential in the above equation. $V_{ext}(r)$ denotes the exterior potentials generated by electron-nucleus interactions, whereas $V_{XC}(r)$ denotes the exchange correlation potentials.

$$\rho(r) = \sum_{i=1}^{N} \Psi_i(r) \Psi_i(r)^*$$

Calculate the quasi particle using the previous equation, where $\Psi_i(r)$ is the electron's single particle wave function.

### 3.9 Exchange and Correlation Functional

The presented DFT analysis is accurate based on the Born-Oppenheimer approximation; however actual use of the DFT approach requires some unrevealed interchange correlation functional. The Pauli Exclusion Principle relates to the exchange part of the exchange and correlation potential, while the columbic potential for interacting electron systems relates to the correlation part. Below are some approximations.

## 3.10 Local Density Approximation

Kohn and Sham introduced the local density approximation as a first approximation for calculating the exchange-correlation functional (LDA). Because they used a hypothetical electron gas system in which the electron travels around a positive charge distribution, the entire system is deemed neutral in this approximation. In this system the number of electrons and the volume of a gas approach infinity, but still the electron density remains constant everywhere. Local density approximations (LDA) produce the best results for non-interacting homogeneous systems with constant electron densities, but LDA does not produce satisfactory results for actual non-homogeneous systems with increasingly unstable electron densities.

$$E_{XC}^{LDA}[n_0] = \int n_0 \in_{XC}^{LDC}(n_0(r))d^3r \quad 3.14$$

The term $\in_{XC}^{LDC} n(r)$ is the homogeneous gas system exchange-correlation energy density at the density of $n(r)$

In the case of electron gas, the system is not constant, LDA is a great accomplishment. This approximation is used when the charge density is not uniform. LDA has a tendency to overestimate binding energies while minimizing ground state energy and ionization energies.

## 3.11 Generalized Gradient Approximation (GGA)

The modified expansion approximation was also unable to correct the situation of real systems more precisely than LDA due to a defect in the exchange correlation functional. As a result, some of its characteristics were lost [10]. for a system with uniform electron density, the local density approximation (LDA) is used, but due to variations in the electric field, many systems possess non- uniform electron density. Before GGA, the local density approximation was employed, but it did not produce helpful findings since it handled electrons locally and did not take into account their interactions. To address this issue, we developed a new method known as generalized gradient approximation (GGA).In this strategy, we leverage exchange correlation, which would be a factor of spin densities as well as their gradient. GGA outperforms the exchange-correlation function for inhomogeneous systems.

$$E_{XC}^{GGA} [\rho(r)] = \int \rho(r) \in_{XC} (\rho(r)\nabla \rho(r))\, dr$$

The most prevalent variant of GGA was introduced by Perdew and Wang in 1992. GGA outperforms IDA in a variety of systems, including charge density variations and the ground state characteristics of light molecules, atoms and solids. We can evaluate the optical properties of predicted stable halides Perovskites (ABX$_3$) using GGA; the most used method in computational chemistry physics. In some circumstances, GGA is unable to provide valid results for some characteristics in these cases; we can apply another technique, which is detailed in the next section. Because of their wide applicability in material science, GGA gives precise findings for all kinds of chemical bonding.

## 3.12 Tran and Blaha Modified Becke-Johnson Exchange Potential (TB-BJ)

In order to determine of electronic structures, Kohn Sham equations are employed in conjunction with GGA or LDA. For various materials, GGA and LDA calculations are limited, However, for insulators and semiconductors, the values of these potentials are vastly underestimated, and the band gap is completely absent in other circumstances. [11].Various potentials, such as the Hubbard potential [12], local density approximation, hybrid functional and dynamic mean field theory [13], have been created to overcome the problem of band gap, however one drawback with this potential is that it is cheap to calculate. Fran and Blaha putBecke Johnson's trade potential to the test [14]. The band gap is characterized by GGA and LDA, which are derived from the Becke Johnson potential, although the band gap of materials is still underestimated. ModifiedBecke Johnson exchange potentials are the result of Tran and Blaha's revisions to the original Becke Johnson exchange potentials. Unlike other computationally intensive methods, this potential (mBJ) accurately calculates a material's band gap.F.Tran and P. Blaha's modified Becke Johnson potential is as follows:

$$v_{x,\sigma}\ mBJ(r) = cv_{x,\sigma}\ BR\ (r) + (3C - 2)\frac{1}{\pi}\sqrt{\frac{5}{12}}\sqrt{\frac{t_0(r)}{n_0(r)}}\ 3.16$$

$v_0\ BR$ Shows the Roussel potential [15], $t_0$ shows the density of electron and "C" is a constant which depend linearly on the average of $\sqrt{\frac{v_n}{n}}$ and $n_o$ indicate the energy density.

$$= \alpha\ +\ \beta\left(\frac{1}{V_{cell}}\int\frac{|\nabla\rho(r')|d^3(r')}{n(r')}\right)^{\frac{1}{2}} 3.17$$

$V_{cell}$ is the unit cell volume and the two free parameters ($\beta = 1.023$ and $\alpha = -0.012$ bohr) are fitted to the experimental band gap. For diverse types of insulators, sp semiconductors, semiconductors, noble gas solids and transition metal oxide, the modified Becke Johnson potential accurately determines the electronic band gaps.

### 3.13 Linearized Augmented Plane Wave Method

The linearized augmented plane wave approach is used to calculate electronic structures and exchange correlation terms using density functional theory. Toanalyze compounds using this technique, first the unit cell split into two parts, one of which is known as non-overlapping atomic sphere, also known as the muffin tin sphere, while other of which is the interstitial section. Anderson predicted the concept of the augmented plane wave technique [16] and used it to the calculation of many common form structures as well as the analysis of various solid properties. The current DFT-based scheme of work can be used to both the thick and unstructured structures of the periodic table. This (LAPW) approach can be used to investigate different phase transitions in a semiconductor material [17]. Slater transformed the augmented plane wave approach [18] into the full potential linearized augmented plane wave approach. In the FP-LAPW technique, potential and charge density are exploited in two separate ways

$$V(r) = \begin{cases} \sum lm\ V_{lm}Y_{lm}(r) & (a) \\ \sum k\ V_k\ e^{ikr} & (b) \end{cases}$$

We used equation (a) to represent the space inside the sphere, which is also known as the muffin-tin sphere, and equation (b) to represent the region outside the muffin-tin sphere (b).

### 3.14 Wien2K Code

The Wien2k programmed which was created in FORTRON and runs on the Linux operating system, uses density functional theory to calculate quantum mechanical electronic structure of solids. The original WIEN 90 code was developed in 1990, and subsequent versions of this code were written in WIEN93, WIEN97, and WIEN2k. It is an electronic system that performs a variety of calculations, including the current GGA, regular TB-mBJ, and no regular TB-mBJ computation, as well as the most recent WIEN2k.The Wien2k simulation tool was used to investigate the properties of the predicted halides perovskite ABX3 inside the context of density functional theory. TheThe FP-LAWP technique is used by Wien2k to estimate the values of various compounds' properties. The Wien2k code is divided into two sections..

I). Initialization

II). Self – consistent field cycle (SCF)

Various Characteristics of compound, such as optimal lattice parameters, volume and bulk modulus and ground state energy are estimated primarily throughout the simulation process by following these two processes. Other features of materials are determined, such as electronic properties such as bonding nature, participation of various states to valence band and conduction, and band profile of materials.

In addition to all of this, optical response and elastic behavior of materials are studied. We investigated and reported the structure and optical behavior of Predicted stable halides Perovskites ABX3 using all of these approaches in this paper.

## 3. Results and Discussion

### 4.1. Structural properties

The predicted stables halides Perovskites materials $ABX_3$ structural properties are investigated in this part using Wien2k, well-known simulation software. The analysis of optimal lattice constants, bulk modulus B, ground state volume $V_O$, pressure derivative and its ground state energy $E_o$ is used to study the structural properties of substances. All of these variables are analyzed as part of the volume optimization process, which stabilizes the entire system. To include exchange and correlation effects in the optimization process, different approximations such as EV-GGA and GGA and are used. This procedure is crucial in the computational analysis of various properties. The energy values of various states are fitted to their associated volume in Birch Murnaghan's equation, which is one of the fundamental equations of state, to generate a volume versus energy parabolic curve, also known as an optimization curve [19]. The energy versus volume curve of $ABX_3$ compounds are shown in Fig. 1. The study of these graphs indicates a general tendency for all chemicals under investigation. The plot shows that when the unit cell energy reduces, its volume increases continually until it achieves its smallest value, which is known as ground state energy, ground state energy corresponding to a particular value of unit cell volume. The unit cell volume grows when the unit cell lattice constant is increased, and for a given value of the volume of the unit cell, the ground state ground state volume ($E_0$) is indicated in Table 4.1.We don't have any experimental or other theoretical data to compare these results with. However, we are confident in the accuracy of our computed results due to the usage of high k-points in the Brillouin zone (BZ), as the same technique of employing a large number of k-points has been documented in the literature.

## 4.2 Electronic Properties

### 4.2.1 Band Structure

The band structure of energy is one of the qualities that distinguish all natural materials from one another. The energy band gap is the best feature for classifying materials into semiconductors, conductors, and insulators. Figure 4.2 depicts the TB-mBJ approximations used to construct the energy band structure of $ABX_3$ predicted stables molecules. The reference point for the Fermi level is zero energy. From the Fig.4.2It is clear that the conduction band minima and valence band maxima of some $ABX_3$predicted compounds lie at different symmetry X and which shows the materials is indirect gap nature, while the valence band maxima and conduction band minima of some compounds like BaCuCl$_3$, ZnLiF$_3$, SrLiF$_3$, ZnLiF$_3$, CsBeCl$_3$ lie at the same symmetry and R at the first Brillion zone, and shows the materials are direct band gap nature. The numerical calculated band gap values for ABX$_3$compound are given in Table 4.2

## 4.2.2 Density of states (DOS)

A material's band structure can also be explained by its density of states. The plot of predicted ABX$_3$compounds' partial density of states (PDOS) and its total density of states (TDOS)is shown in Fig.4.3. To determine the contributions of electronic states to charges carriers around Fermi level using TB-mBJ functional PDOS and TDOS calculations. The Fermi level was set at the maximum of the valence band in our calculation, and the states with the highest contribution were shown for clarity.

While for dealing BaAgBr$_3$ it is clear that the valence band the band near -5.7eV to 0eV originated Ba-s and Ag-d orbital. The lower portion of the conduction bands is caused by Ag-s, while the upper part is caused by Br-d orbital. The majority contribution in the valence band is Ba-s, Ba-d and Ag-d orbital and conduction band the majority contributed is Br-d and Ba-d orbital. In BaAgCl$_3$the valence band near to -6eV to 0eV originated Ag-d and Cl-p orbital while the Cl-p orbital the majority contribution is start from -6eV to -3eV while for Ag-d is started -2.5eV to 0eV. In conduction band the majority contribution is Ba-f and Ba-d orbital while small contribution of Ag-s.In valence band of BaCuCl$_3$ the majority contribution are Cu-d ,Cl-p and Cl-d orbital. The majority contribution of Cu-d is started from -1.5eV to 0eV.while the contribution of Cl-p and Cl-d is high at -3.5ev to -7eV. In the conduction band of BaCuCl3 the majority contribution is Ba-f orbital while contribution of Cl-p is small. In the valence band of CaAgI$_3$ the majority contribution are I-p, Ag-s and Ag-p at the energy range of -5.5eV to -3.5eV. While in this region contribution of I-p is highest. In conduction band the majority contribution are I-d, Ag-p and Ag-s but contribution of I-p is highest.

In valence band of CaCuBr$_3$ the majority contribution are Cu-d and Br-p orbital, the contribution of Cu-d is highest at -1.9eV to 0eV while Br-p contribution is high at -7.5eV to -3.4eV. while in conduction band the majority contribution is Ca-d and Br-d orbital but highest contribution is Ca-d at 1eV to 0.1eV.In valence band of CaCuI$_3$ the majority contribution are I-p and Cu-d orbital, but I-p contribution is highest in the range of -8eV to -2eV. While in conduction band the majority contribution is Ca-d at 2.8eV to 0.3eV but there is little contribution of I-d and I-f orbital. In the valence band of CeBeCl$_3$ the majority contribution are Cs-p and Cl-p the contribution Cs-p is high at -4.5eV to -5.5eV white contribution of Cl-p is high at -5.5eV to 0eV.while in conduction band the majority contribution are Cs-f orbital the orbital of Be-p are contributed in both valence band and conduction band but Be-p contribution is high in conduction band. In the valence band of CsRhCl$_3$ the majority contribution are Cl-p, Rh-d and Cs-p orbital. The contribution of Cl-p is started from -2.1eV to -6eV.while Rh-d orbital contributed both in conduction band and valence band and cross the Fermi level, which shows that the compound is metallic in nature. While in conduction band the majority contribution is Cs-d and Cl-d orbital.

In the valence band of CsRuCl$_3$ the majority contribution is Cs-p orbital and Cl-p orbital clear from the graph orbital of Cs-p are distributed in small energy range while orbital Cl-p are distributed in the energy range of -6eV to 0eV, the orbital of Ru-d are present in both conduction band and valence and cross the Fermi level give prediction of metallic nature. While in conduction band majority contribution are Cs-f and Cl-p orbital. The DOS plot shows that valence band of CsTcI$_3$ the majority contribution is Cs-p in small energy gap while Tc-d and I-p orbital give high contribution in the energy range of -4eV to -1eV.while in conduction band majority contribution is T-d in the high energy range. The compounds of KIrI$_3$ the valence band consist of majority contribution of I-p in enery range of -6.1eV to 0eV and Ir-d orbital contributed both valence and conduction and cross Fermi level which show that compound of CsTcI3 is metallic in nature.

The valence band of KNiI$_3$ consist of majority contribution of Ni-d and I-p orbital the high contribution is Ni-p orbital clear from the fig 4.2.Both orbital contributed valence and conduction band at the Fermi level mean valence and conduction band overlap there is negligible energy gap, so the martial is metallic in nature. While in conduction majority distribution are K-d orbital in energy of 5.5eV to 15eV.The DOS plot of RbIrBr$_3$ show the material is metallic in nature because of crossing of Fermi level. While in conduction band the majority contribution are Rb-d orbital. The DOS plot of RbPdBr$_3$ show the material is metallic in nature because of crossing of Fermi level. While in conduction band the majority contribution are Rb-d orbital while little contribution of Rb-p and Pd-p orbital. The DOS plot of RbRhBr$_3$ show the material is metallic in nature because of crossing of Fermi level. While in conduction band the majority contribution is Rb-d orbital while little contribution of Rb-d orbital.

In the valence band of SrLiF$_3$consist of majority contribution of Li-p and F-p orbital in the energy range of -2.1eV to 0eV.while in conduction band the majority contribution is Sr-d orbital. In the valence band of TlBeF$_3$ the majority contribution is Tl-d orbital and F-d orbital clear from the graph orbital of Tl-d are distributed in small energy range while orbital F-p are distributed in the energy range of -8eV to 4eV. while in conduction band majority contribution are Be-s and Be-p orbital. while small contribution of F-d orbital. In the compound of TlSrF$_3$ the valence band consist of majority contribution of Tl-s and F-p in the energy of -2eV to 0eV, while conduction band the majority contributions are Tl-p in energy of 6.5eV to 9.5eV, while Sr-d distributed in the of 10.8ev to 16eV.while contribution of Tl-d is very small. In the valence band of ZnLiF$_3$ the majority contributions are Zn-d, Li-p, Li-s and F-d. The contribution of Zn-d orbital is in the small energy range. While Li-p orbital is in the energy range of -3eV to 0eV. While in conduction band the majority contribution is Zn-s. The DOS of ZnScF$_3$ shows valence and conduction band electronic states distribution, it is show that valence band consist of Zn-d and F-p majority distribution.

### 4.3 Optical Properties

The study of optical characteristics of materials is essential because it explains how compounds react to electromagnetic waves. In the context of the Density Functional Theory (DFT), all of the optical properties of halides Perovskites ABX$_3$ are computed using the FP-LAPW technique. Below is a detailed description of optical constants like the dielectric function, which includes imaginary parts and real, reflectivity, refractive index, optical conductivity and absorption coefficient.

### 4.3.1 Dielectric Function

A detailed explanation of a material's response to an incident electromagnetic disturbance is the dielectric function. It consists of both real and imaginary parts, and represented by following equation.

$$\varepsilon(\omega) = \varepsilon_1(\omega) + i\varepsilon_2(\omega)$$

Where $\varepsilon_1(\omega)$ show real part and $\varepsilon_2(\omega)$ shows the imaginary part of the complex dielectric function. The imaginary component of the dielectric is an essential quantity that explains all transitions between energy states, from conduction to valence bands.

Dielectric function's real component $\varepsilon_1(\omega)$ indicates the polarizability of a compound after its electric distortions and defines the refractive index of a material. While the spectrum of optical absorption is directly proportional to the imaginary part $\varepsilon_2(\omega)$ of the dielectric constant. It also describes the absorption edge, a remarkable optical characteristic. Furthermore, it determines information about a solid's electronic band structure. Because the imaginary component of the dielectric function describes the absorption behavior of any material. The absorption of a transparent compound, such as glass is lower. The dielectric function imaginary part will be low due to poor absorption; just the real part is left. This real part is now expressed in simple real number. As a result, the real parts of dielectric functions comprise the majority of data in the low absorption zone. When there is a lot of absorption, and however, the dielectric function's complicated character must be recognized. For selected compounds, the real portion $\varepsilon_1(\omega)$ of the dielectric function was determined using Kramer's-Kronig dispersion relation.

The predicted ABX$_3$ compounds dielectric function real parts are shown in the Fig. 4.4.1. The BaAgBr3 the static dielectric function is gotten as $\varepsilon_1(0) = 3.5$. It begins to increase from its zero frequency limit and reaches its peak of 7.5 at 5eV. After reaching another peak at 17.5eV For the range of 18eV to 20eV, it begins to decline farther beneath zero in the negative scale. This material exhibits metallic behavior when the real dielectric function $\varepsilon_1(\omega)$ is negative, otherwise it is dielectric. Similarly the static dielectric function $\varepsilon_1(0)$ is obtained for BaAgCl$_3$, BaCuCl$_3$, CaAgI$_3$, CaCuBr$_3$, CaCuI$_3$, CsRhCl$_3$, CsRuCl$_3$, CsBeCl$_3$, PdRhBr$_3$, SrLiF$_3$, TlBeF$_3$ are 3, 3.5, 5, 4.5, 6, 8, 4, 3.2, 8, 1.8, 2.8 respectively. The highest static dielectric function $\varepsilon_1(0)$ is obtained for KNiI$_3$, KIrI$_3$ and CsTcI$_3$ are 13, 14 and 16 respectively. This indicates slightly higher polarizability. For some compound like BaAgCl$_3$, BaAgBr$_3$, BaCuCl$_3$, CaCuBr$_3$, CsRuCl$_3$, CsBeCl$_3$, CaCuI$_3$, SrLiF$_3$, TlBeF$_3$ ZnScF$_3$, TlSrF$_3$ the peak value of real dielectric function $\varepsilon_1(\omega)$ are observed in UV region, and for CaAgI$_3$ the peak value is observed in visible region. While RbIrBr$_3$, KIrI$_3$, KNiI$_3$, the peaks value is observed in infrared region.

Similarly, imaginary part of dielectric function for ABX$_3$ has been computed in Fig 4.4.2 The threshold values of dielectric functions are occurring at 0.2eV, 0.5eV, 0.6eV for CaCuBr$_3$, CsCuCl$_3$, and CaAgI$_3$ respectively while for BaCuCl$_3$, BaAgBr$_3$, BaAgCl$_3$, CsBeCl$_3$ the threshold value occurring 2eV, 2.1eV, 2.4eV, 2.7eV respectively while high threshold value are observed for compound TlBeF$_3$, ZnLiF$_3$, TlSrF$_3$ and SrLiF$_3$ are 4.9eV, 5.4eV 6.4eV and 9.1eV respectively. The absorption peaks of BaAgCl$_3$, BaCuCl$_3$, CaCuBr$_3$, CsRuCl$_3$, and CsBeCl$_3$ are found in the energy range of 5eV-20eV. While absorption maxima for ZnLiF$_3$ and TlSrF$_3$ are found in the energy range of 6eV to 12eV. These absorption peaks are caused by the interband transition area. The Complex dielectric function contains all of the required data for determining additional optical parameters like as refractive index, reflectivity, transmission coefficient, extinction coefficient, adsorption coefficient, transmissivity among others.

### 4.3.2 Complex Refractive Index

A compound's refractive index is an optical property that explains how light behaves in that compound. It is determined by the frequency of the incident light beam. Refractive index is also consisting of real and imaginary part mean complex in nature. It is stated as;

$$\boldsymbol{\eta(\omega) = \mathrm{n}(\omega) + k(\omega)}$$

Where n (ω) is real part of complex refractive index and $k(\omega)$ is the imaginary part of complex refractive index.

The complex refractive index n(ω) real part is similar as the ordinary refractive index, whereas the imaginary part k(ω) of complex refractive index is known as the extinction coefficient. Shown in figure 4.5. For predicted stables halides Perovskites $ABX_3$, it imaginary and real part of the complex refractive index is presented. The value of static function n(0) at zero frequency (static refractive index) for $KNiI_3$ are 3.5 for RBIrBr3 are 3 for $RbRhBr_3$ and $CsRhCl_3$ are 2.9 for $CsBeCl_3$ are 2.8 for $CaCuI_3$ are 2.5 for $CaAgI_3$ are 2.3 for $RbPdBr_3$ and $CaCuBr_3$ are 2.1 for $CsRuCl_3$ are 2 for $BaAgBr_3$ and $BaCuCl_3$ are 1.9 for $BaAgCl_3$ are 1.7 for $TlSrF_3$ are 1.5 while for $ZnLiF_3$ are 1.25.

From the Figure 4.5.1 we observed that all compounds refractive index values rises from static frequency limit and achieve a number of peaks in different energy ranges. The highest peak are observed for $BaCuCl_3$ and $CsBeCl_3$ are 7.5eV for $TlSrF_3$ are 7.2eV for $BaAgCl_3$eV are 6.5eV for $BaAgBr_3$ are 5eV for $CsRuCl_3$ and $BaAgBr_3$ are 5eV for $KNiI_3$ are 4.8eV for $CaCuBr_3$ are 5eV for $CaCuI_3$ are 2.75eV for $CaAgI_3$ are 2.5eV while for $RbIrBr_3$ and $RbPdBr_3$ are 2eV. While compound of $ZnLiF_3$ it refractive index is constant up to the energy range of 25eV, then decrease. The compound of $SrLiF_3$ gives maximum peak at 24eV then decrease the value of refractive index below unity as energy increase up to 27.5eV. It is clear that all maximum peaks are observed in visible and ultraviolet region.

The optical energy absorbed in the optical medium during light transmission is described by the extinction coefficient k (w), it has been demonstrated that the dielectric function and refractive index real parts follow similar trends. While the dielectric function and refractive index imaginary component has comparable characteristics.

In figure 4.5.2 It can seen that some compounds like $BaAgBr_3$, $BaAgCl_3$, $BaCuCl_3$, $CaAgI_3$, $CaCuBr_3$, $CaCuI_3$, $CsRuCl_3$, $CsBeCl_3$, $ZnLiF_3$ has significant value in the ultraviolet region. While compound like $KIrI_3$, KNiI3, $RbIrBr_3$, $RbPdBr_3$, and $ZnScF_3$ has significant value both visible and ultraviolet region. There these compounds are very important for optoelectronic devices.

### 4.3.3 Absorption Coefficient

The absorption coefficient α (ω) is a compound's linear attenuation coefficient when it reacts with light [38]. The dielectric function is crucial in determining a compound's absorptioncoefficient. The absorption coefficientsPredictedhalidePerovskites ABX3 are given in figure 4.5

At various high symmetry points, the peaks are associated to the inter band transition on the electronic band spectrum. In the energy range of 0 to 40eV, all compounds show various peaks., but the highest absorption peaks observed among all is at 26eV for SrLiF$_3$ and the second highest peaks are observed for CaAgI$_3$, CaCuBr$_3$ and CACuI$_3$ at 28eV, while for BaAgCl$_3$ and BaCuCl$_3$ are observed at 18eV. The broad absorption energy range of these compounds promises uses in the optoelectronic as well as optical instrument operating energy range.i.e.3 -40eV[39].

### 4.3.4. Optical Conductivity

To explore the spectrum of optical conductivity for predicted stables halides Perovskites ABX$_3$compoundsthe optical conductivityσ(ω) is an importance parameter to it,asshown in figure 4.7

The electronic conductions related to the incidence of electromagnetic radiations are represented by this spectrum. Figure shows those compounds' highest peaks σ(ω) arises in the visible to ultraviolet range, (1.7eV to 30eV).

The maximum optical conductivity is observed for SrLiF$_3$ at 24eV while the second maximum peaks are observed for BaAgBr$_3$, BaAgCl$_3$ and BaCuCl$_3$ at 18eV. Scissor operator is also used for better optical conductivity σ(ω).

### 4.3.5 Reflectivity

The optical reflection of light from a compound's surface is significant. The complex refractive index can be used to calculate the optical reflectivity of any material. We in Fig 4.8.For the understudy compounds, maximum reflectivity was discovered in the low to high energy region. Among all these compounds highest reflectivity is observed for BaAgCl$_3$, BaAgCl$_3$ at 18eV and for BaCuCl$_3$ at 22.5eV.

**Conclusions**

The structure, electrical structure, and optical responses of the simple cubic Perovskites ABX$_3$ were calculated using DFT in Wien2k code. To account for exchange and correlation effects, the generalized gradient approximation (GGA) and TB-mBJ were used. The conclusions are as follow:

From Structure calculation we have evaluated that all predicted halides Perovskites ABX$_3$ are simple cubic structure and are stables. The band gaps of these entire compounds are calculated. Band gap of some compound like SrLiF$_3$, TlSrF$_3$, ZnLiF$_3$, TlBeF$_3$are 9.1eV, 6.4 eV, 5.4 eV, 4.9 eV respectively which shows that these compounds are insulator in nature. While the band gap of compound likeBaAgCl$_3$, BaAgBr$_3$, BaAgCl$_3$ and CsBeCl$_3$ are 2eV, 2.1 eV, 2.4eV, and 2.7eV respectively, and shows that these compounds are semiconductor in nature. There is no band gap found in some compound like KIrI$_3$, KNiI$_3$, CaCuI$_3$, RbPdBr$_3$, RbIrBr$_3$, and RbRhBr$_3$. Shows that these compounds are metallic in nature. The band gap is also calculated by PDOS and TDOS. Furthermore, it was discovered that some compounds had a straight band gap, making them appropriate for photovoltaic applications. While some compounds have indirect band gap found, so these indirect band gap materials are used as absorber in solar cells. Optical measures such as complex dielectric function, reflectivity, and absorption coefficient, and conductivity, extinction coefficient have been used to investigate the optical response of these compounds. The highest static dielectric function $\varepsilon_1(0)$ is obtained for KNiI$_3$, KIrI$_3$ and CsTcI$_3$ are 13, 14 and 16 respectively. This shows that the polarizability is slightly higher. Among all these compounds highest reflectivity is observed for BaAgCl$_3$, BaAgCl$_3$ at 18eV and for BaCuCl$_3$ at 22.5eV.

# References

1. Lang, L., Yang, J. H., Liu, H. R., Xiang, H. J., & Gong, X. G. **(2014)**. First-principles study on the electronic and optical properties of cubic ABX$_3$ halide perovskites. *Physics Letters A*, *378*(3), 290-293.

**2.** Körbel, S., Marques, M. A., &Botti, S. **(2018).** Stable hybrid organic–inorganic halide perovskites for photovoltaics from ab initio high-throughput calculations. *Journal of Materials Chemistry A*, *6*(15), 6463-6475.

3. Li, C., Lu, X., Ding, W., Feng, L., Gao, Y., &Guo, Z. **(2008).** Formability of ABX$_3$ (X= F, Cl, Br, I) halide perovskites. *ActaCrystallographica Section B: Structural Science*, *64*(6), 702-707.

4. Yi, T.; Chen, W.; Cheng, L.; Bayliss, R. D.; Lin, F.; Plews, M. R.; Nordlund, D.; Doeff, M. M.; Persson, K. A.; Cabana, J. Investigating the Intercalation Chemistry of Alkali Ions in Fluoride Perovskites. Chem. Mater. **2017**, 29, 1561−1568.

5. Zhang, Y.; Lin, J. D.; Vijayaragavan, V.; Bhakoo, K. K.; Tan, T. T. Y. Tuning Sub-10 Nm Single-Phase NaMnF3 Nanocrystals as Ultrasensitive Hosts for Pure Intense Fluorescence and Excellent T$_1$ Magnetic Resonance Imaging. Chem. Commun. **2012**, 48, 10322− 10324.

6. Demartin, F.; Campostrini, I.; Castellano, C.; Russo, M. Parascandolaite, KMgF$_3$, a New Perovskite-Type Fluoride from Vesuvius. Phys. Chem. Miner. **2014**, 41, 403−407.

7. Li, C.H., Lu, X.G., Ding, W.Z., Feng, L.M., Gao,Y.H., Guo, Z.M. Formability of ABX$_3$(X = F, Cl, Br, I) halide perovskites, ActaCrystallogr. 64 (6) **(2010)** 702–707.

8. Liu, X., Fu, J., & Chen, G. **(2020)**. First-principles calculations of electronic structure and optical and elastic properties of the novel ABX3-type LaWN$_3$perovskite structure. *RSC Advances*, *10*(29), 17317-17326.

9. Becke, A.D; Roussel, M.R. *J. Phy.Rev*. **1989**, 39, 3761- 3767.

10. Winner, E; Krakauer, H; Weinert, M; Freeman, A. *J. Phys.Rev*.1981, 24864-875.

11. Perdew, J.P; Wang, Y.J. *Phy.Rev*. B 45, 13244- 13249.

12. Slater, J.C,J. Phys. Rev, **1937**, 51, 846-851.

13. Slater, J. C. *J.Adv. Quantum Chem*. **1964**, 1, 35- 58

14. Schwarz, K; Blaha, P. *J. Comput. Mat. Sci.* **2003**, 28, 259.
15. Birch, F. (1947). Finite elastic strain of cubic crystals. *Physical review*, *71*(11), 809.
16. Murtaza, G., Ahmad, I., Maqbool, M., Aliabad, H. R., &Afaq, A. **(2011).** Structural and optoelectronic properties of cubic CsPbF3 for novel applications. *Chinese Physics Letters*, *28*(11), 117803.
17. Perdew, J. P., Burke, K., &Ernzerhof, M. **(1996).** Generalized gradient approximation made simple. *Physical review letters*, *77*(18), 3865.
18. Becke, A. D. **(1988).** Density-functional exchange-energy approximation with correct asymptotic behavior. *Physical review A*, *38*(6), 3098.
19. Peralta, G.; J. I., &Bokhimi, X. (2020). Discovering new perovskites with artificial intelligence. *Journal of Solid State Chemistry*, *285*, 121253.

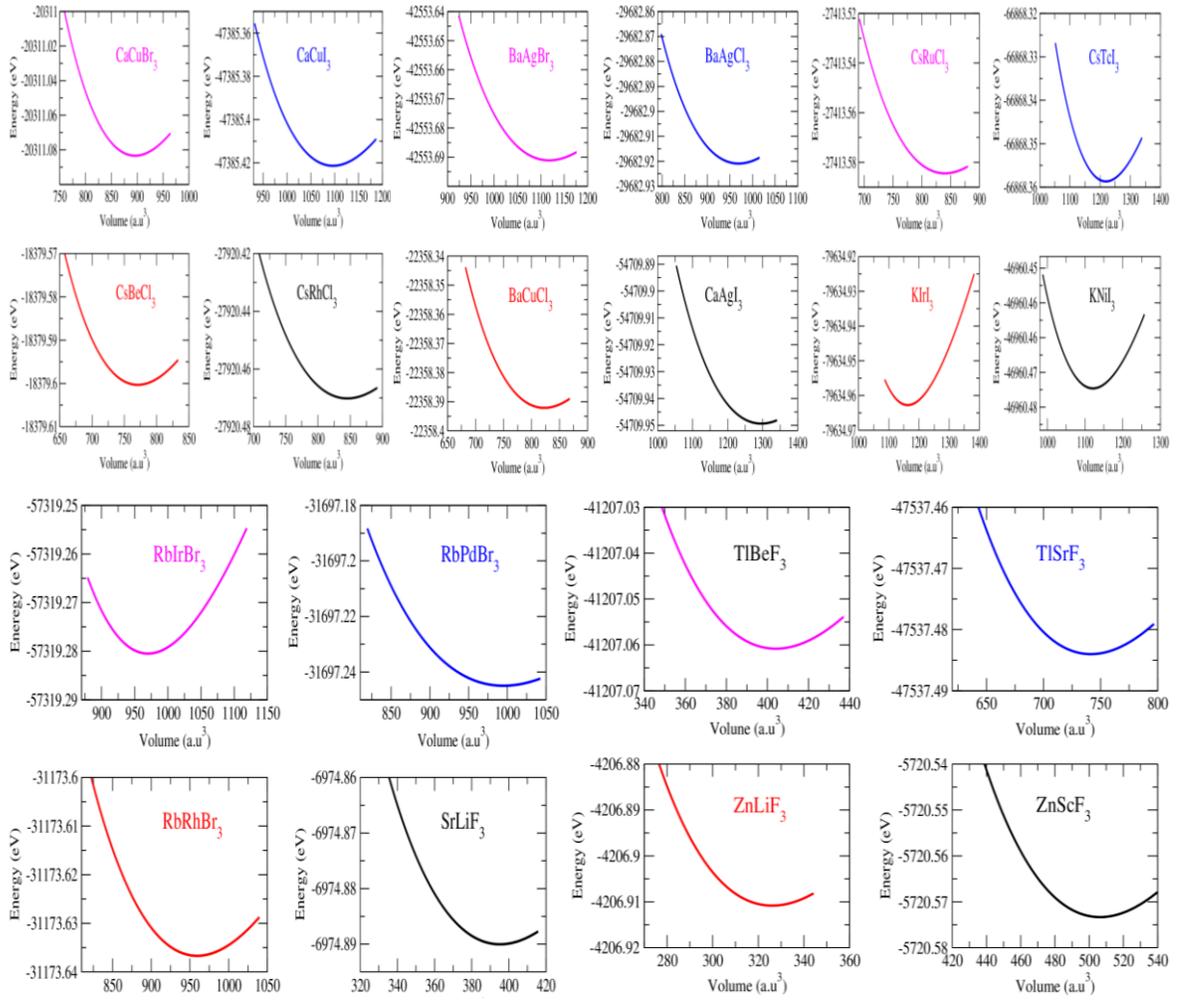

**Fig. 1. Unit cell volume optimization curve of compounds stable halides ABX₃ perovskites**

**Table 1** Calculated value of structure parameter i.e. Ground state energy ($E_0$) Derivatives of Bulk Modulus (BP), Ground States Volume ($V_0$) for predicted stables Halides Perovskites of $ABX_3$

| | Lattice constant ($a_0$) | Bulk Modulus B (GPa) | Bulk Modulus Derivative (BP) | Optimized Volume ($V_0$) | Ground Energy ($E_0$) |
|---|---|---|---|---|---|
| $CaAgI_3$ | 5.768<br>5.619[a] | 25.387 | 5.0 | 1295.975 | -5479.949 |
| $CsTcI_3$ | 5.653<br>5.616[a] | 30.479 | 5.0 | 1219.265 | -66868.359 |
| $KIrI_3$ | 5.564<br>5.650[a] | 37.872 | 5.0 | 1162.658 | -79634.963 |
| $KNiI_3$ | 5.497<br>5.609[a] | 28.541 | 5.0 | 1128.541 | -46960.473 |
| $BaAgBr_3$ | 5.491<br>5.379[a] | 30.086 | 5.0 | 1117.622 | -42553.691 |
| $CaCuI_3$ | 5.457<br>5.395[a] | 57.389 | 5.0 | 1096.478 | -47385.421 |
| $RbPdBr_3$ | 5.283<br>5.167[a] | 36.583 | 5.0 | 995.049 | -31697.245 |
| $RbIrBr_3$ | 5.238<br>5.364[a] | 44.365 | 5.0 | 970.055 | -57319.280 |
| $BaAgCl_3$ | 5.236<br>5.122[a] | 34.422 | 5.0 | 969.214 | -29682.921 |
| $RbRhBr_3$ | 5.219<br>5.162[a] | 40.834 | 5.0 | 959.272 | -31173.637 |
| $CaCuB_3$ | 5.101<br>5.034[a] | 84.551 | 5.0 | 895.906 | -20311.083 |
| $CsRhCl_3$ | 5.004<br>4.905[a] | 47.235 | 5.0 | 845.484 | -27920.470 |
| $CsRuCl_3$ | 4.993<br>4.883[a] | 47.494 | 5.0 | 840.016 | -27413.584 |
| $BaCuCl_3$ | 4.958<br>4.861[a] | 40.979 | 5.0 | 822.591 | -22358.392 |
| $CsBeCl_3$ | 4.852<br>4.795[a] | 39.197 | 5.0 | 770.885 | -18379.600 |
| $TlSrF_3$ | 4.790<br>4.724[a] | 40.703 | 5.0 | 741.523 | -47537.484 |
| ZnScF$_3$ | 4.218<br>4.159[a] | 82.089 | 5.0 | 506.403 | -5720.573 |
| TlBeF$_3$ | 3.9123<br>3.867[a] | 88.919 | 5.0 | 404.097 | -41207.061 |
| SrLiF$_3$ | 3.883<br>3.803[a] | 70.912 | 5.0 | 395.239 | -6974.890 |
| ZnLiF$_3$ | 3.642<br>3.571[a] | 86.977 | 5.0 | 326.087 | -4206.911 |

[a]Ref. [40]

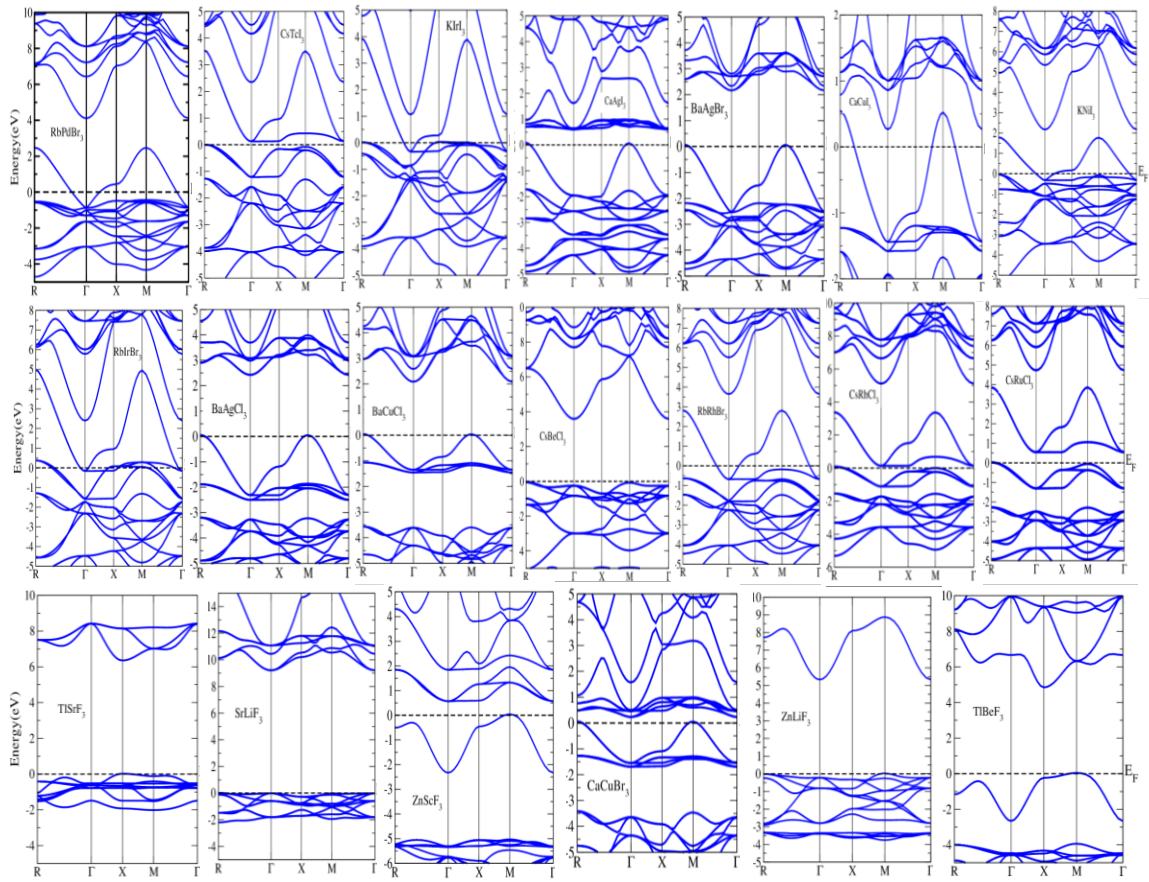

**Figure 4.2.** Band profiles of predicted halides Perovskites ABX$_3$

**Table 2.** Band gap value of Predicted Stables halides Perovskites ABX$_3$

| Compounds | Band Gap E$_g$ ( in eV) | Experimental work |
|---|---|---|
| **CaAgI$_3$** | 0.6 | Null |
| **CsTcI$_3$** | 0 | Null |
| **KIrI$_3$** | 0 | Null |
| **KNiI$_3$** | 0 | Null |
| **BaAgBr$_3$** | 2.1 | Null |
| **CaCuI$_3$** | 0 | Null |
| **RbPdBr$_3$** | 0 | Null |
| **RbIrBr$_3$** | 0 | Null |
| **BaAgCl$_3$** | 2.4 | Null |
| **RbRhBr$_3$** | 0 | Null |
| **CaCuBr$_3$** | 0.2 | Null |
| **CsRhCl$_3$** | 0 | Null |
| **CsRuCl$_3$** | 0.5 | Null |
| **BaCuCl$_3$** | 2 | Null |
| **CsBeCl$_3$** | 2.7 | Null |
| **TlSrF$_3$** | 6.4 | Null |
| **ZnScF$_3$** | 0.5 | Null |
| **TlBeF$_3$** | 4.9 | Null |
| **SrLiF$_3$** | 9.1 | Null |
| **ZnLiF$_3$** | 5.4 | Null |

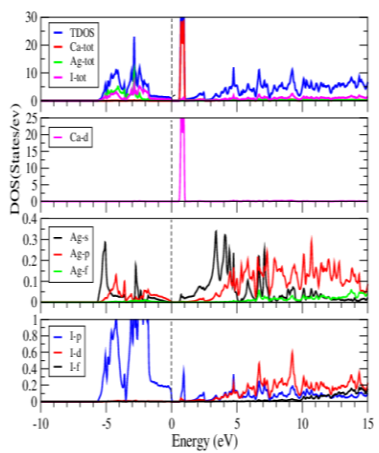
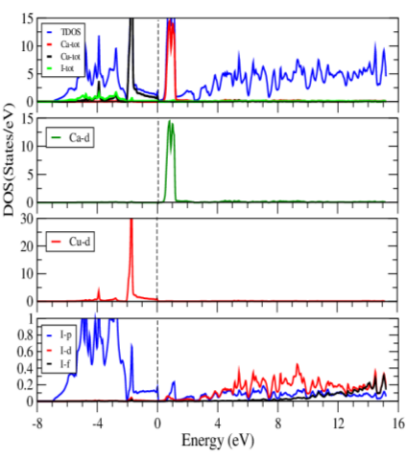
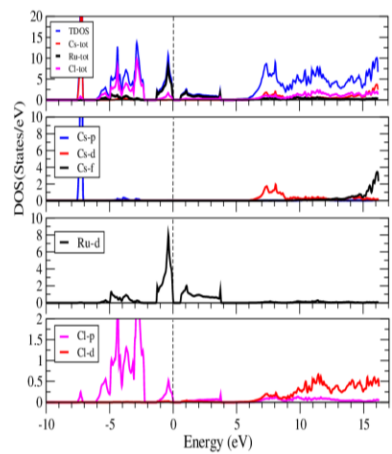

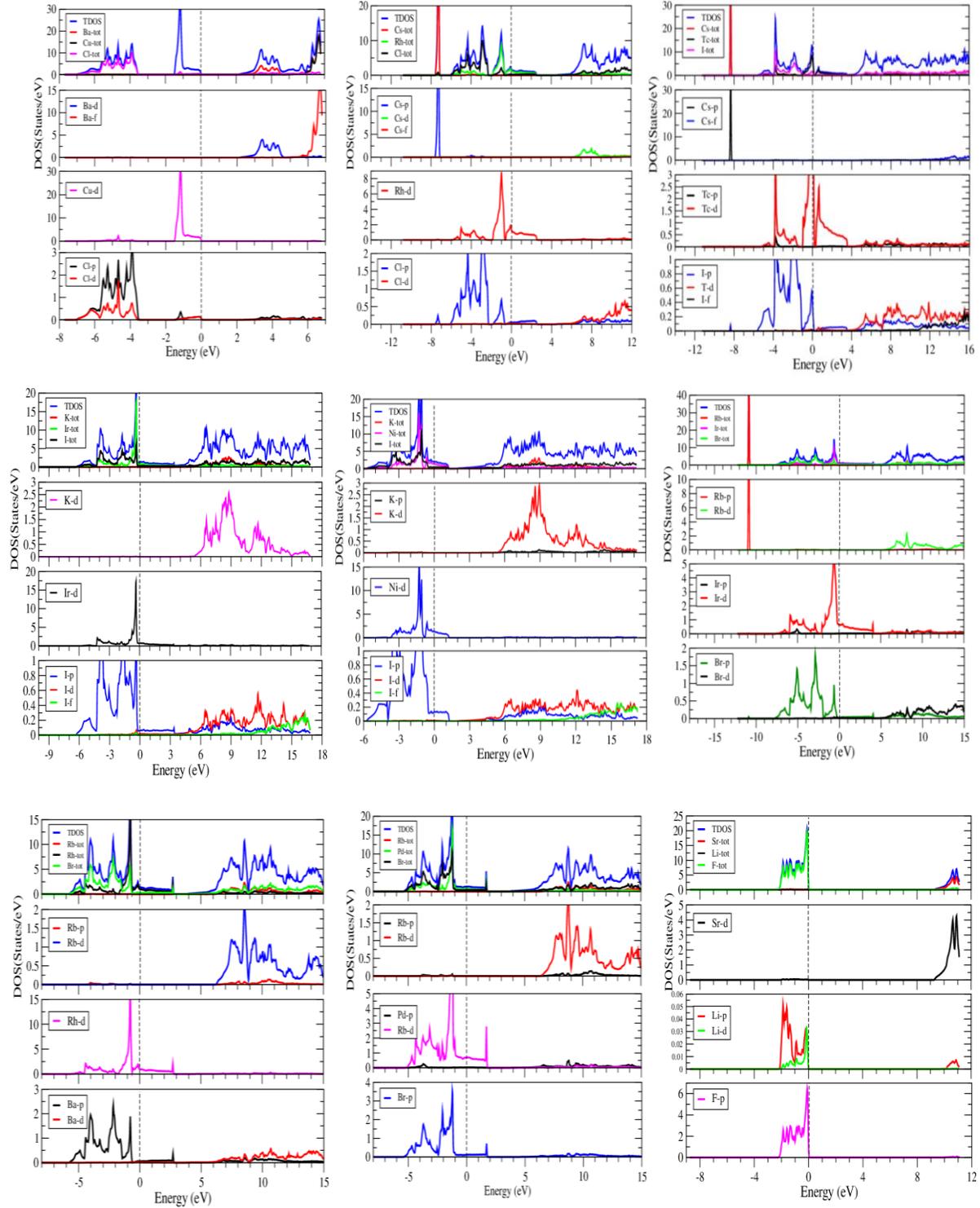

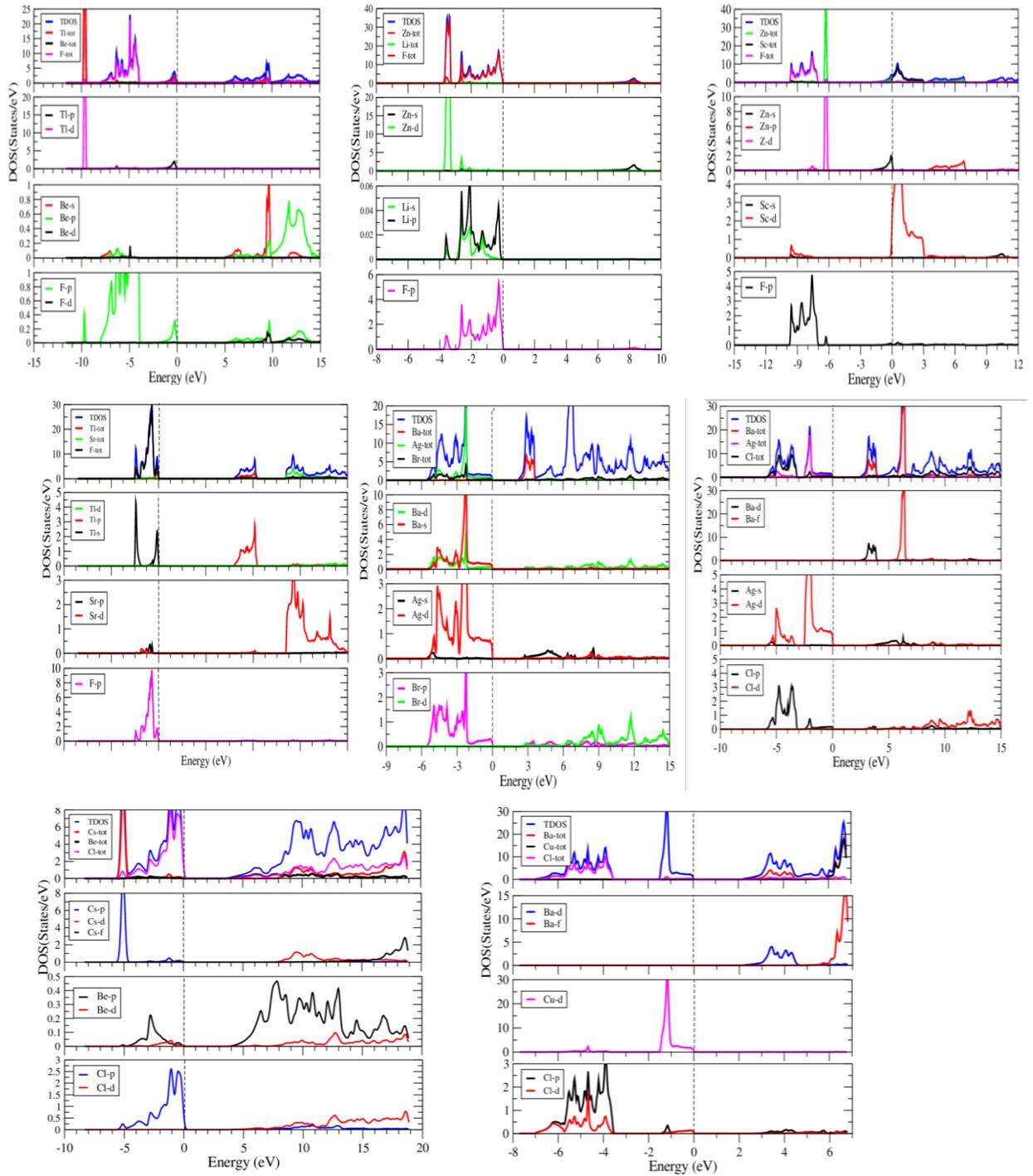

**Fig 4.3. TDOS and PDOS for predicted ABX₃ Perovskites**

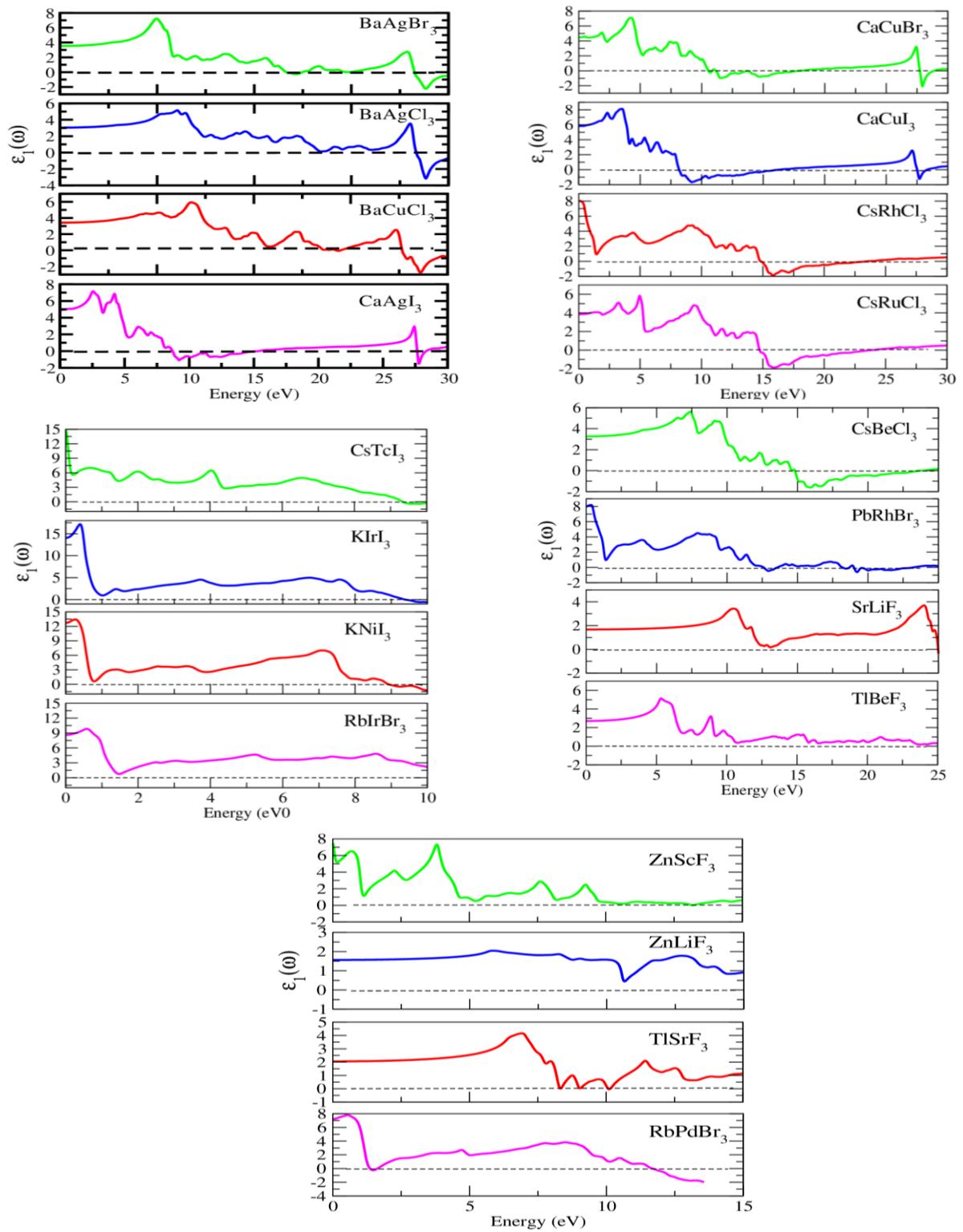

**Fig 4.4.1. Real Part of Dielectric function for Predicted halides Perovskites ABX$_3$**

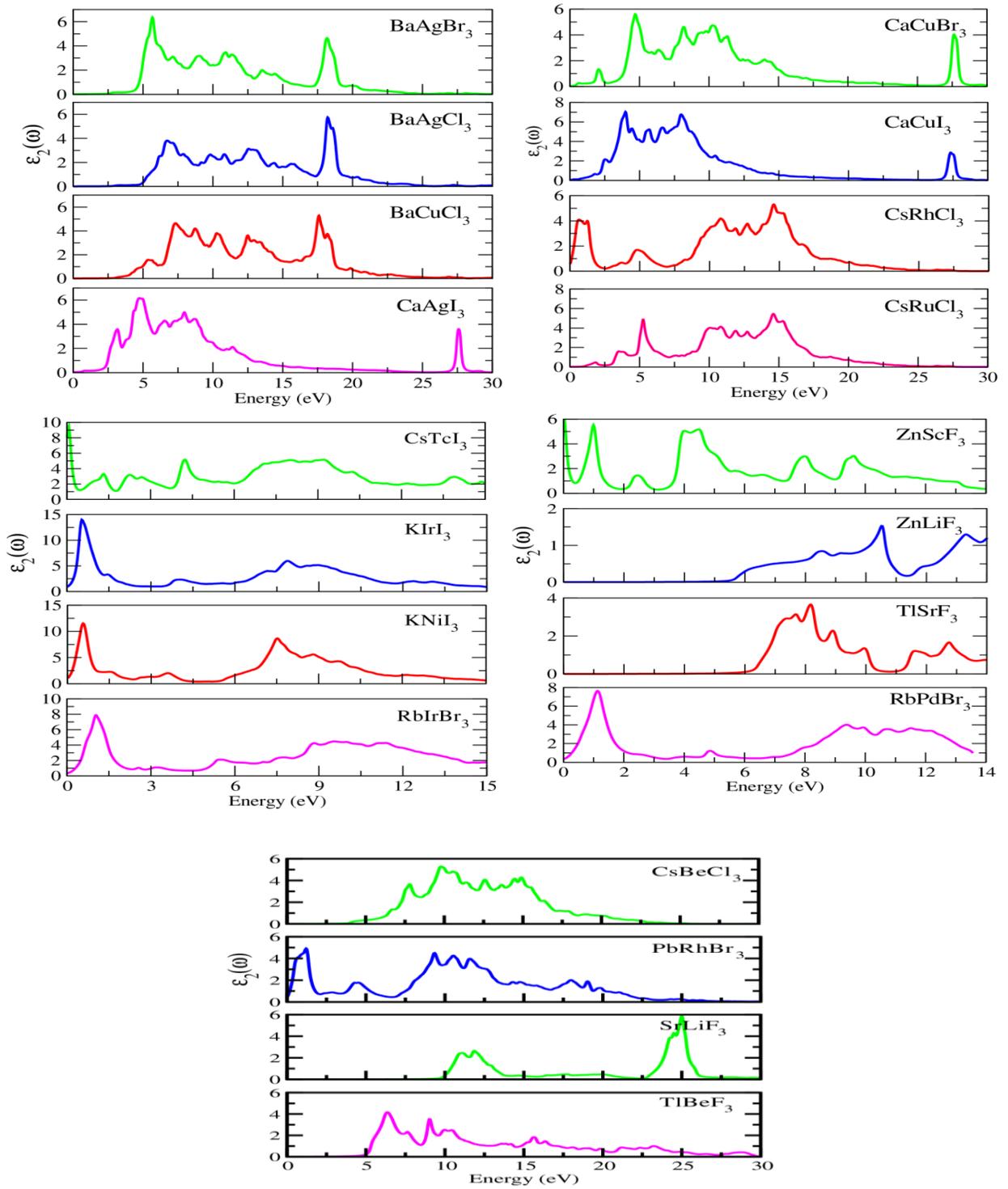

**Figure 4.4.2. Imaginary part of dielectric function for predicted halides Perovskites ABX3**

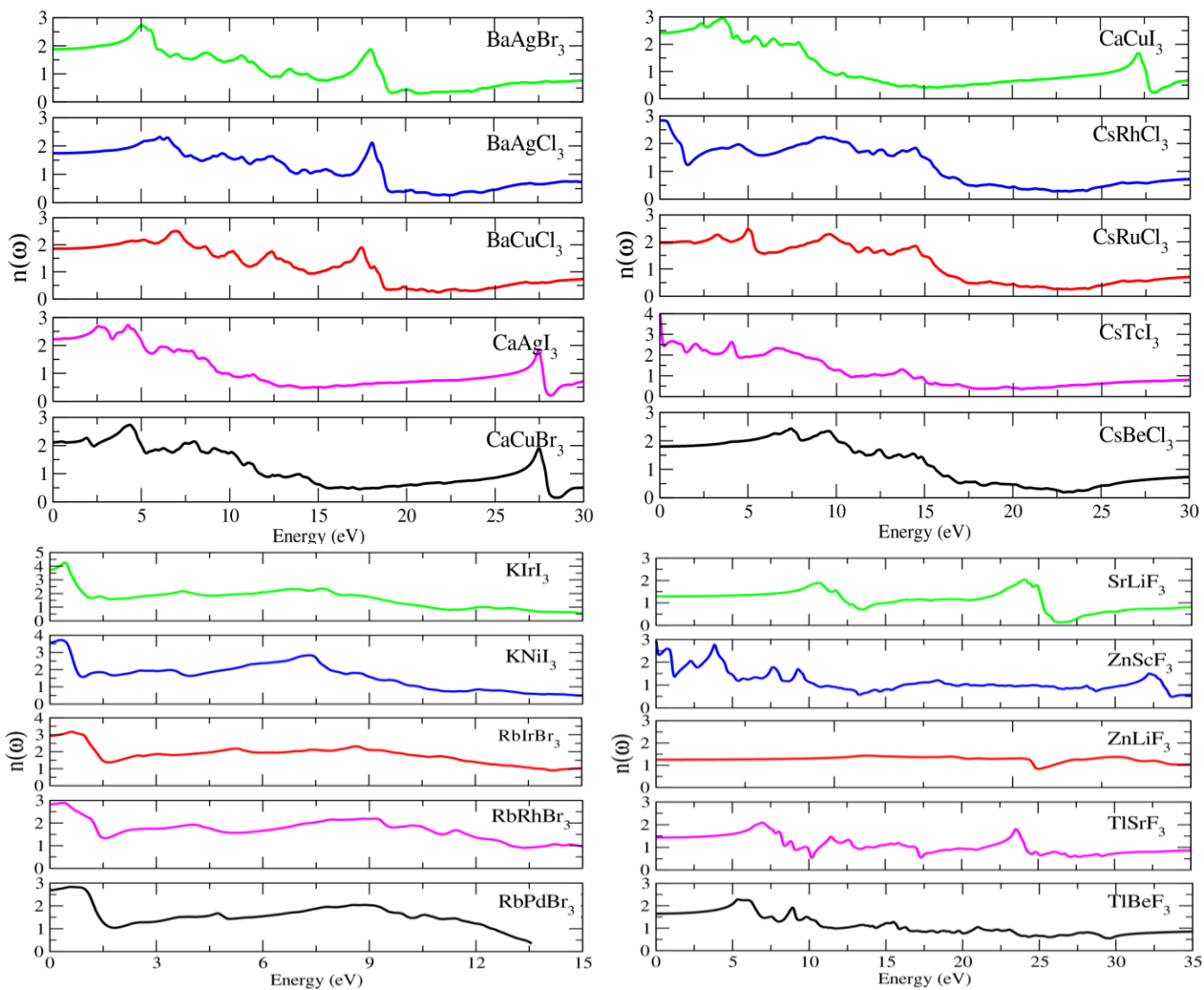

Figure 4.5.1. Real part of complex Refractive index for predicted halides Perovskites ABX$_3$

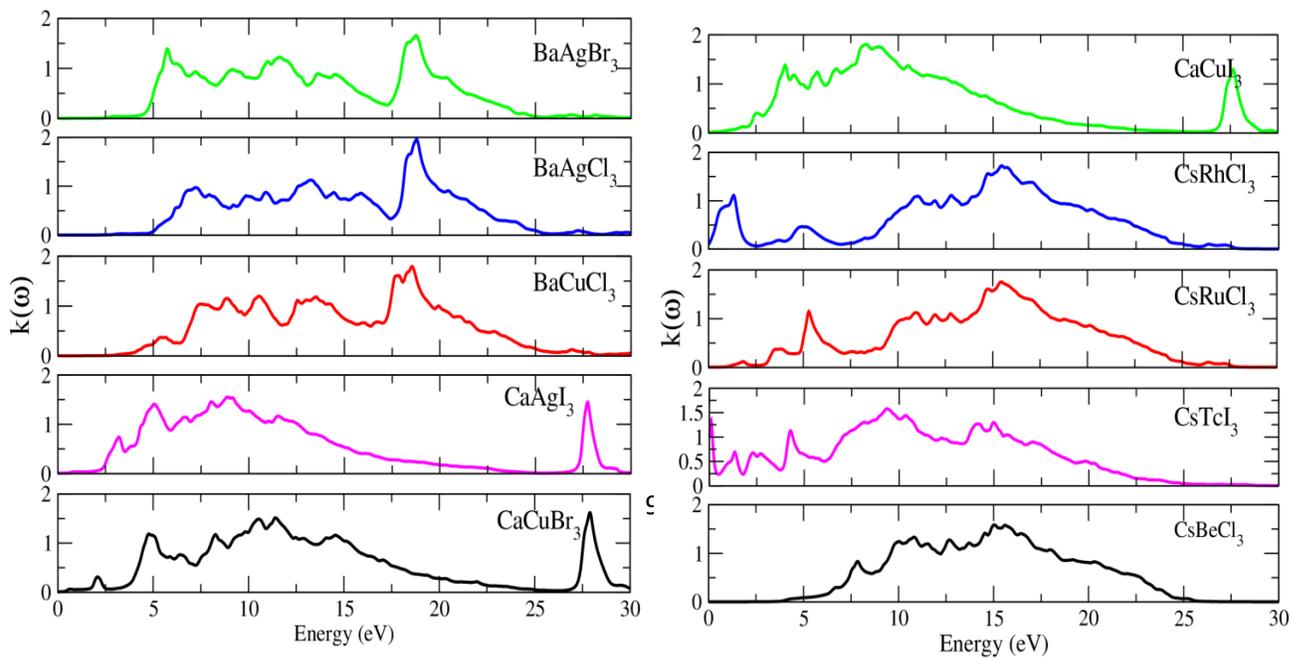

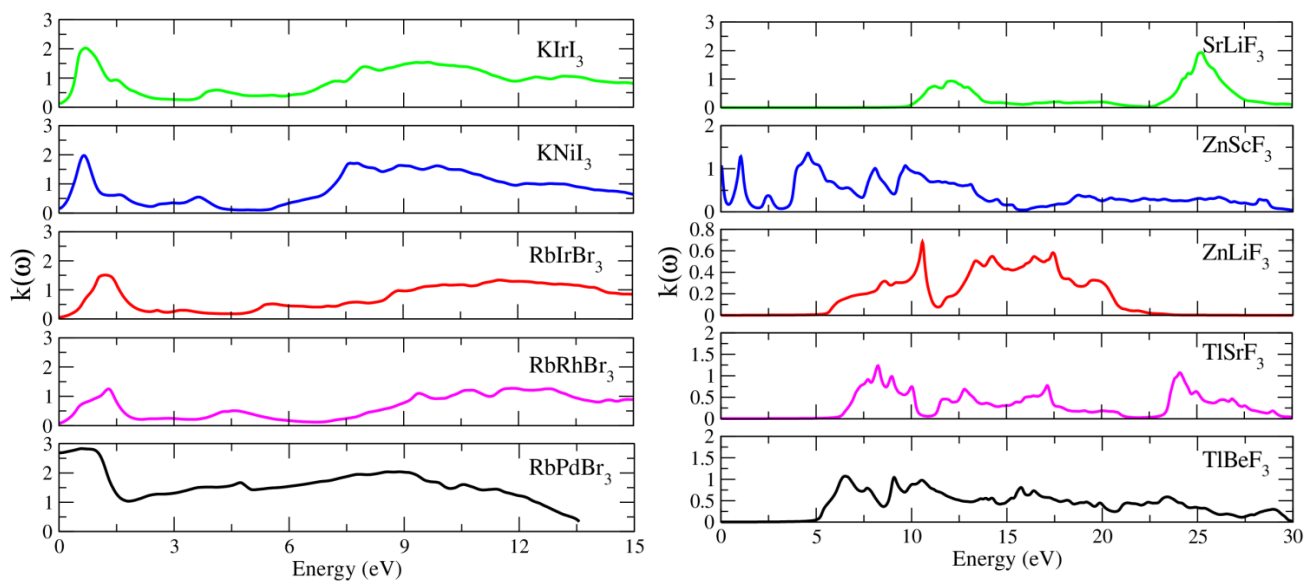

**Figure 4.5.2. Imaginary part complex reactive index for predicted halides Perovskites ABX3**

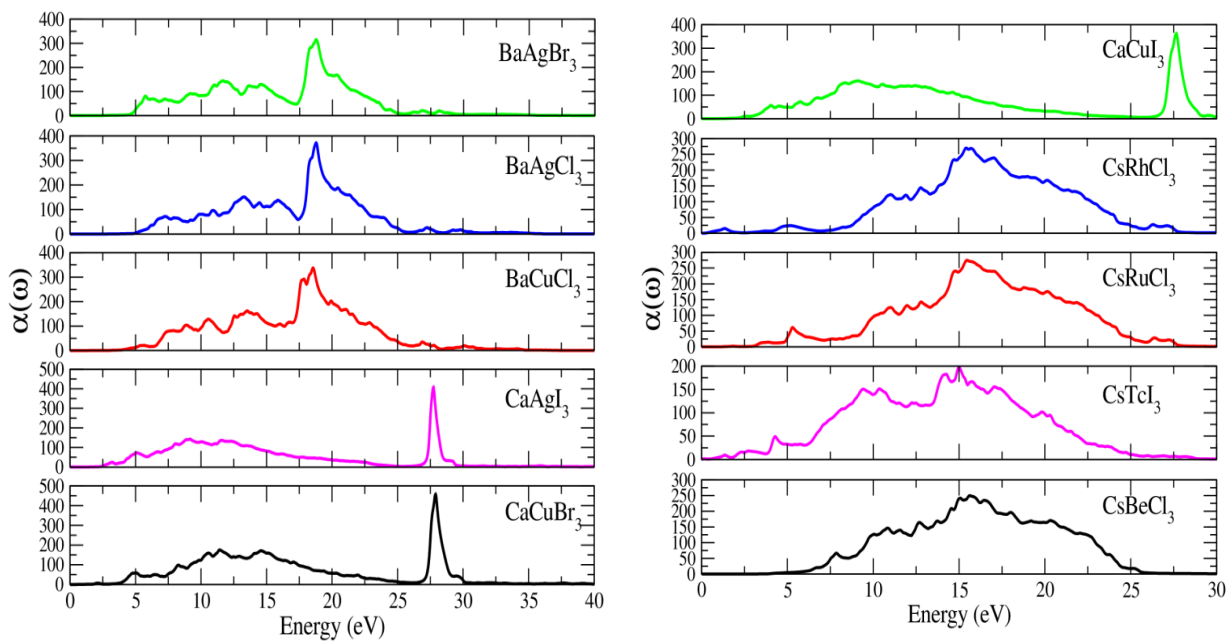

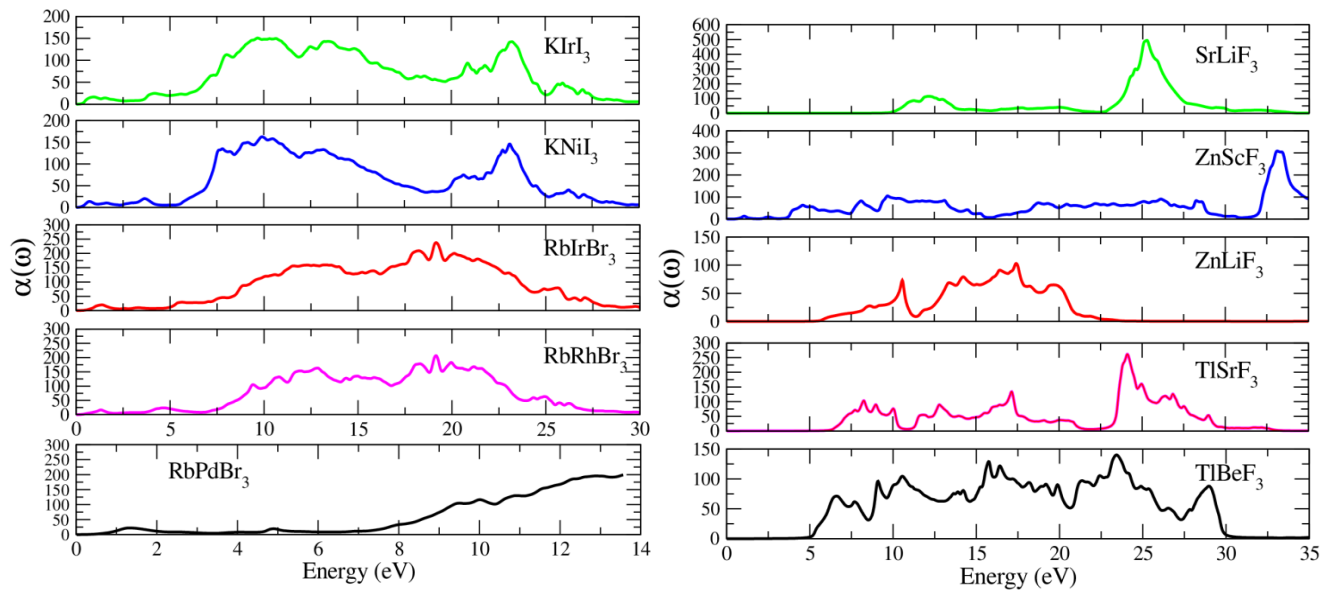

**Figure 4.6. Absorption Coefficient for predicted halides perovskites ABX₃**

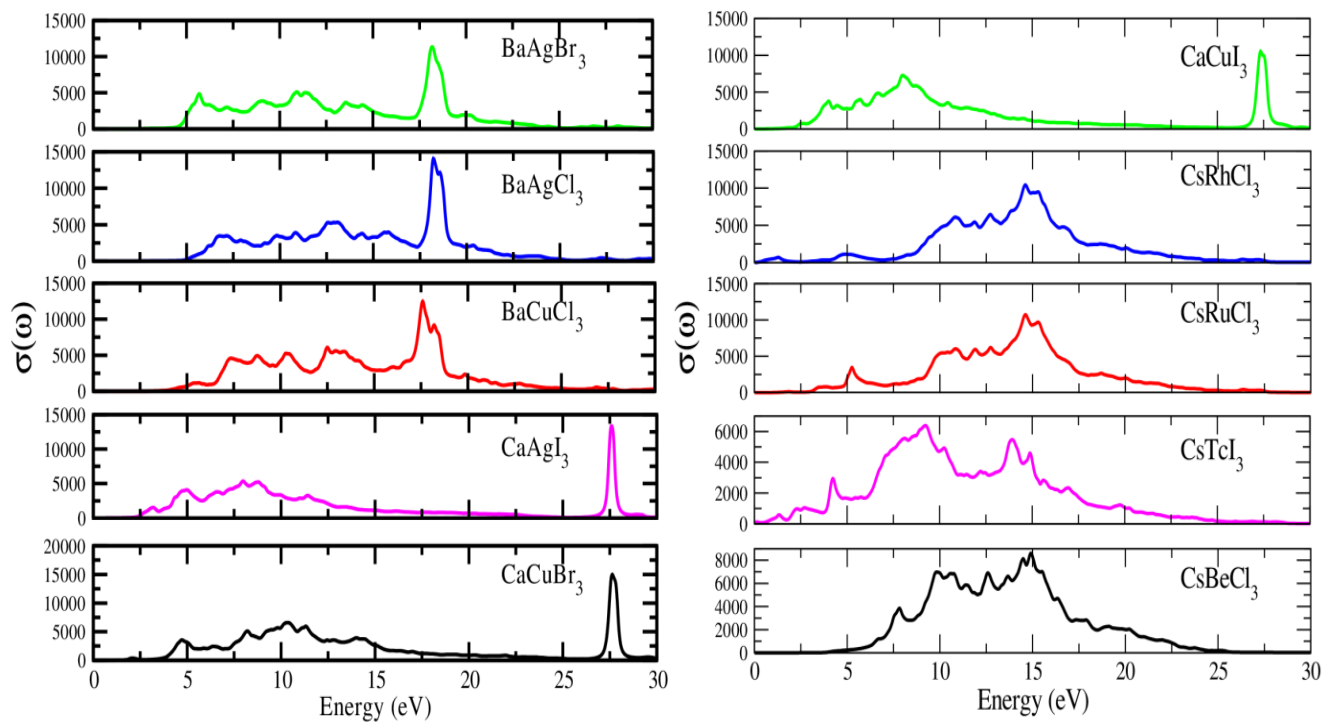

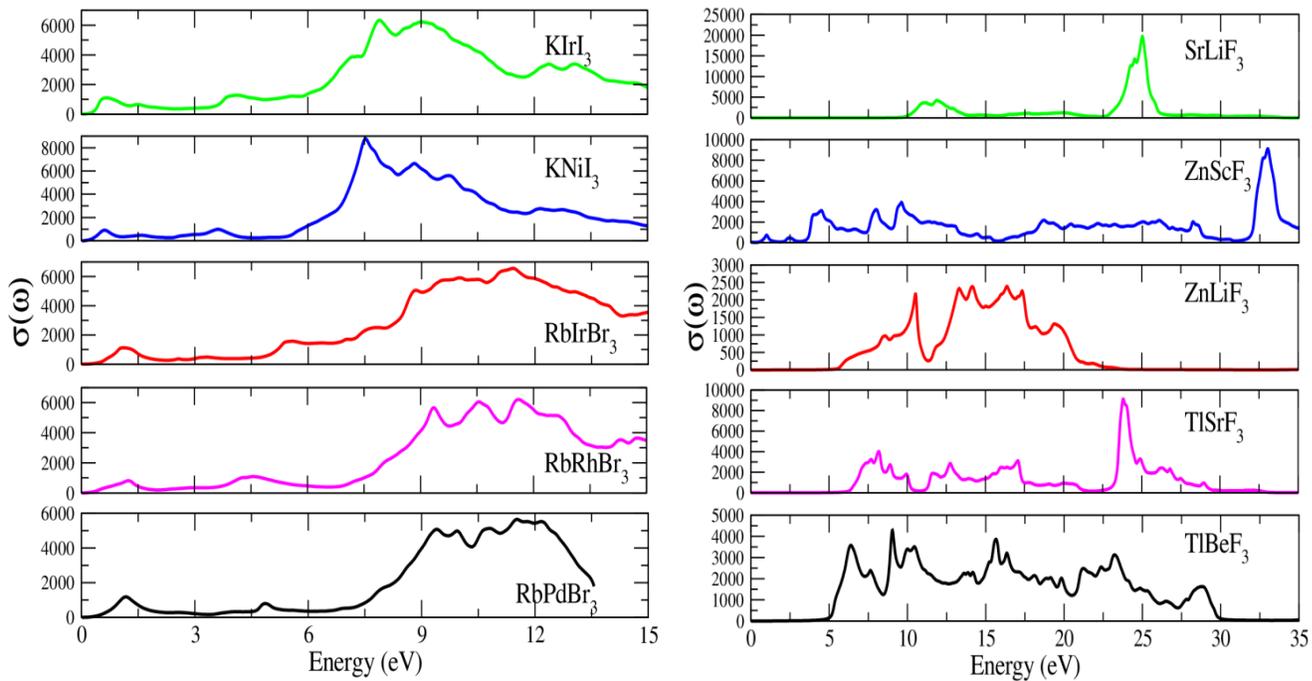

**Figure 4.7. Optical conductivity for predicted halides Perovskites ABX$_3$**

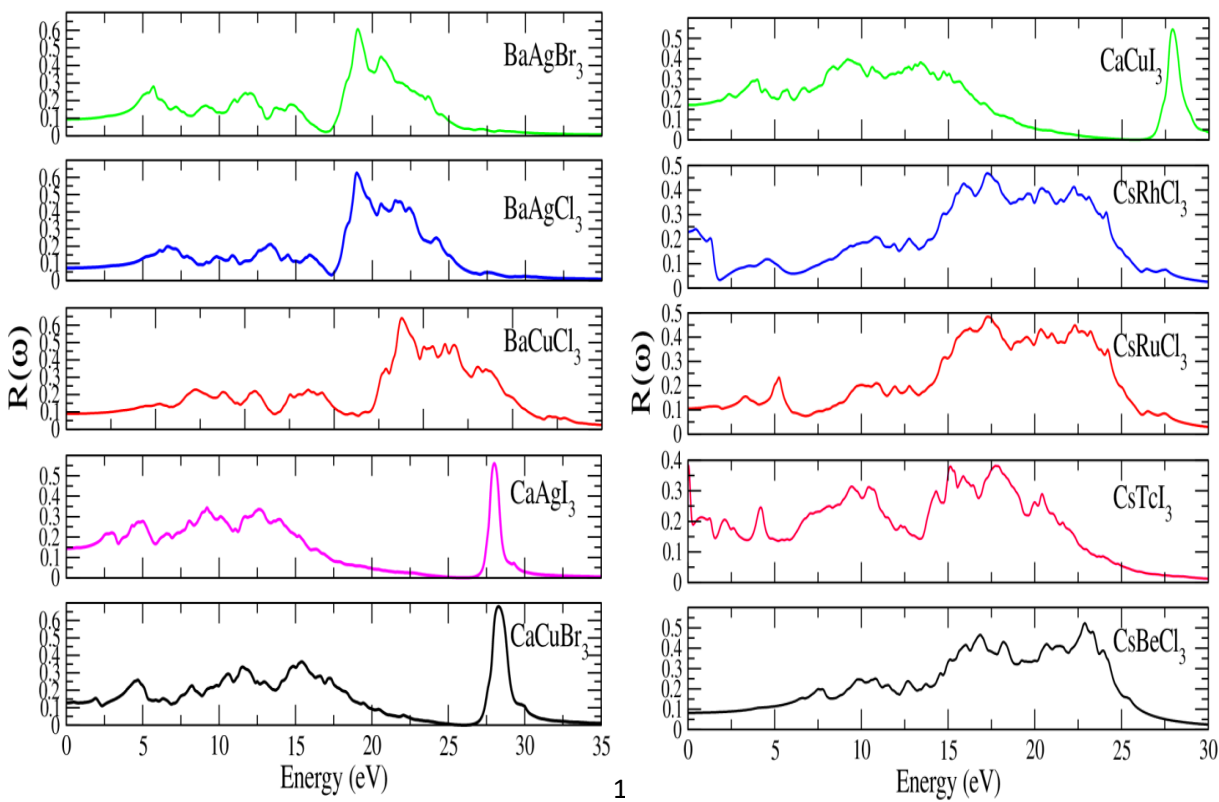

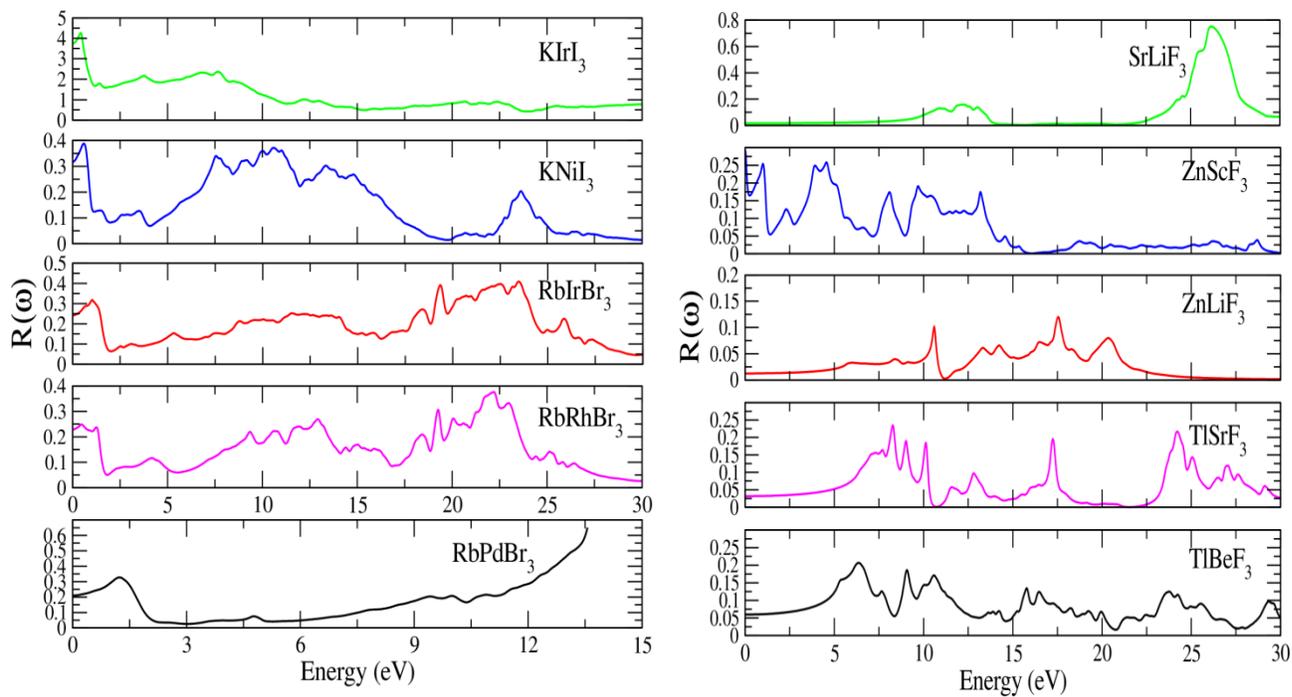

**Figure.4.8. Evaluate Reflectivity for predicted halides perovskites ABX₃**